\documentstyle[preprint,aps,epsf]{revtex}
\newcommand{\adots} {{\mathinner{\mkern2mu\raise1pt\hbox{.}\mkern2mu
\raise4pt\hbox{.}\mkern2mu\raise7pt\hbox{.}\mkern1mu}}}

\begin{document}
\title{Bosons Doubling}

\author{Herv\'e Mohrbach}
\address{LPLI-Institut de Physique, 1 blvd D.Arago, F-57070 Metz, France}

\author{Alain B\'{e}rard}
\address{LPLI-Institut de Physique, 1 blvd D.Arago, F-57070 Metz, France}

\author{Pierre Gosselin}
\address{Universit\'e Grenoble I, Institut Fourier, UMR 5582 CNRS-UJF, \\
UFR de Math\'ematiques, \\
BP74, 38402 Saint Martin d'H\`eres, Cedex, France}
\date{\today}
\maketitle
\begin{abstract}
It is shown that next-nearest-neighbor interactions may lead to unusual
paramagnetic or ferromagnetic phases which physical content is radically
different from the standard phases. Actually there are several particles
described by the same quantum field in a manner similar to the species
doubling of the lattice fermions. We prove the renormalizability of the
theory at the one loop level.
\end{abstract}

\section{Introduction}
The most practical way to build a physically relevant quantum field theory
starts with a choice of a suitable Lagrangian. More than the Hamiltonian
formalism, the Lagrangian enforces both the Lorentz invariance and the
symmetry principles. Yet the difficulty is the choice of the terms to
include in the Lagrangian allowed by the symmetries. This choice has been
dictated for a long time by the principle of renormalizability. Solely
renormalizable quantum field models were considered as sensible physical
theories in particle physics. The need of infinitely many coupling constants
to cancel the UV divergences generated in the framework of the perturbation
expansion made the non-renormalizable theories be rejected. Today any
realistic quantum field is considered as an effective theory valuable only
in a given range of energy. In this context, renormalizability. is no more
considered as a fundamental physical requirement. It may be possible that
the quantum field theories we are familiar with, are low energy
approximations of a theory that may not even be a field theory.

Any effective theory includes both renormalizable and non-renormalizable
interactions. But the characterization of the second ones was modified by
looking into their importance at low energy. There, we expect they are
highly suppressed. So non-renormalizable interactions may be excluded from
the start because their influence on the dynamics decreases with the
physical energy scale; i.e., they do not change the universality class of
the model, but their influences grow when we consider the high energy
dynamics. These non-renormalizable interactions are then interpreted as the
influence of some degrees of freedom relevant at higher energy on the low
energy physics. For instance, the heavy particle perturbative elimination
leads to an effective Lagrangian for the light particles containing an
infinite number of non-renormalizable interactions expressed in terms of the
light degrees of freedom.

Consider as an example a single scalar component Lagrangian with higher
derivative interactions of the form $\varphi \Box ^{n}\varphi $. We follow
the discussion of Steven Weinberg's book \cite{weinberg}. Such a term make a
contribution of the form $(q^{2})^{n}$ to the free propagator. Thus it would
not have the simple pole expected but $n$ such poles usually at complex
value of $q^{2}$ \cite{kuti}. They could be interpreted as particles with
negative norm which violate unitarity \cite{lee}. Following Weinberg's
argument if this non-renormalizable operator has a coefficient of order $%
M^{-2(n-1)}$ ($M\gg m$), then the extra poles are at $q^{2}$ of order $M^{2}$
and we can not neglect all the other non-renormalizable interactions. In
other words the higher derivative terms are non-renormalizable interactions
generated by the elimination of a particle of mass $M$. It is then difficult
to describe the physics beyond the heavy particle threshold without this
particle as a dynamical degree of freedom. To due this job without the true
degrees of freedom we precisely need the fine tuning of an infinity of
non-renormalizable interactions. Therefor the truncation of the effective
Lagrangian to terms of the form $\varphi \Box ^{n}\varphi $ is a poor
approximation.

In this letter we consider a one component scalar field theory with higher
derivative interactions regularized on a lattice. Usually, a lattice or a
continuum theory differ only by non-renormalizable interactions leading to
the same low energy physics. But it is well known that in some cases,
singular configurations or topological defects at the scale of the cutoff
may appear on the lattice, polluting the numerical simulation. Following the
previous discussion the usual attitude is to suppress such configurations by
improving the action \cite{hasenfratz} in order to recover the same
continuum limit than in renormalized perturbation theory. In the present
letter we consider a model with a next-nearest-neighbor interaction and
choose to vary the dynamics close to the cutoff scale to look if this may
change the physics at large distance. Then at first sight, it seems that the
preceding reasoning would forbid such a point of vue: the true degrees of
freedom or all non-renormalizable interactions are needed at the scale of
the cutoff. This is true if we our theory is an effective low energy theory.
But if it appears to be renormalizable, our model is correct at all energy
scales and we don't need to include all other non-renormalizable
interactions in the action.

Keep in mind that the classification of the interactions is usually done by
simple dimensional analysis (power counting arguments in perturbation
theory) but that non-perturbative study may invalid this classification. For
example, the formation of small positronium bound states (of the size of the
cutoff) in strong massless QED, generates by taking into account the
anomalous dimensions, new relevant (renormalizable) operators. The
condensate of these bound states breaks the chiral symmetry. The IR feature
of the resulting vacuum are thus modified compared to the perturbative one 
\cite{bar},\cite{miranski}.

Our goal is to look if continuum physics exists beyond the class of
traditionally renormalizable theories. By considering our theory on a
lattice we find a free propagator containing many minima (in euclidean
space). These ones are similar to the doubling fermions and will be
interpreted as different particles, some of them having different masses. By
a particular fine tuning of the coupling constants, we find a renormalizable
theory at least at the one loop level. We also introduce a 16-components
field $\Phi _{\alpha },$ in such a manner that each component is responsible
of the excitations around one minima. In this manner each of these
excitations is now interpreted as low energy excitations of different
fields. This formalism allows us compute the one loop effective potential.

The present letter is a generalization of a preceding work published in two
papers \cite{af1} where the accent was put on the breakdown of the
Poincar\'{e} symmetry by the second pole of a propagator containing two
minima, leading to an antiferromagnetic vacuum. The lattice is a good
regulator since contrary to the other ones it regularizes the quantum
fluctuations as well as the saddle point. This may be important if
non-homogeneous saddle point are present. In the present paper we will work
with a trivial vacuum, the generalization to a ferromagnetic one being
trivial.

\section{The model}

We consider the following single component scalar field action in a d
dimensional lattice: 
\begin{eqnarray}
S\left[ \varphi (x)\right] &=&\sum_{x}\left\{ -\frac{1}{2}\varphi (x)\left[
A\varphi (x)+\sum_{\mu }\left( J\varphi \left( x+e_{\mu }\right) +K\varphi
\left( x+2e_{\mu }\right) \right) \right] \right\}  \nonumber \\
&&+\sum_{x}(\frac{\widetilde{m}^{2}}{2}\varphi (x)^{2}+\frac{\lambda }{4!}%
\varphi (x)^{4})  \label{action}
\end{eqnarray}
where the coefficients $A$, $J$, $K$ are chosen to be positive. The theory
describes a paramagnetic (P) or a ferromagnetic (F) phase. A negative sign
for $J$ leads to an antiferromagnetic phase with the breaking of the Lorentz
invariance. We don't want to discuss such a situation here. It is well known
from renormalization group argument that next-nearest-neighbor ferromagnetic
coupling are irrelevant for the description of the P or F phase at least
near the phase transition. In particle physics language those operators have
a decreasing influence on the dynamics as we move away from the UV scaling
regime towards the physical energy scales, in other words they do not change
the universality class of the model. In a P or F phase the important modes
are the modes near zero. In particular they are responsible for the
instability leading to a phase transition. It will be shown below that for
this model all the relevant modes lie in fact around each edge of the
Brillouin zone. These fast fluctuating modes are then relevant as precursor
of a phase transition to an antiferromagnetic phase. We aim to study the
influence of these modes in the continuum limit. As usual, the fluctuations
around a minimum of the propagator are interpreted as particle like
excitations. So we will show that our model describes the dynamics of $2^{d}$
particles. This is similar to the fermion doubling on the lattice except
that our particles are not degenerates.

\section{The elementary excitations}

The particles in the mean-field approximation are given by the free
propagator: 
\begin{equation}
G^{-1}\left( p\right) =-A+\widetilde{m}^{2}-2\left( J\sum_{\mu }\cos p_{\mu
}+K\sum_{\mu }\cos 2p_{\mu }\right)  \label{prop}
\end{equation}
which has the particularity to have $2^{d}$ minima in each edge of the
Brillouin zone if: 
\begin{equation}
K>\frac{J}{d}  \label{condi}
\end{equation}
It's advantageous to divide the Brillouin zone 
\begin{equation}
\mathcal{B}=\left\{ p_{\mu },\left| p_{\mu } \right|\le\pi \right\} ,
\end{equation}
into $2^{d}$ restricted zones, 
\begin{equation}
\mathcal{B}_{\alpha }=\left\{ \left| p_{\mu }-P_{\mu }\left( \alpha \right)
\right| \le \frac{\pi }{2}\right\}
\end{equation}
whose centers are at 
\begin{equation}
P_{\mu }\left( \alpha \right) =\pi n_{\mu }\left( \alpha \right) ,
\end{equation}
where $n_{\mu }\left( \alpha \right) =0,1$ and the index $1\le\alpha
\le 2^{d}$ is given by 
\begin{equation}
\alpha =1+\sum_{\mu =1}^{d}n_{\mu }\left( \alpha \right) 2^{\mu -1}.
\end{equation}
The propagator for the zone $\mathcal{B}_{\alpha }$ is $G_{\alpha
}^{-1}\left( q\right) =G^{-1}(P\left( \alpha \right) +q).$ It turns out that
all the Brillouin zones $\alpha =1..2^{d},$ contain particle like
excitations.

In particular: 
\begin{equation}
G_{1}^{-1}\left( 0\right) =-A-2d(J+K)+\widetilde{m}^{2}
\end{equation}
So we chose arbitrary: $A=-2d(J+K).$ If we assume that the mass term is
finite, the free propagator in the limit $a\rightarrow 0$ is (with the
lattice spacing explicitly reintroduced) : 
\begin{equation}
G_{\alpha }^{-1}\left( p\right) =Z(\alpha )p^{2}+m^{2}(\alpha )+O(a^{2}p^{4})
\end{equation}
with the mass given by: 
\begin{equation}
m^{2}(\alpha )=m^{2}+\frac{Jd}{a^{2}}\sum_{\mu =1}^{d}n_{\mu }(\alpha )
\end{equation}
where $m^{2}=\frac{\widetilde{m}^{2}}{a^{2}}$. The condition of finiteness
of the mass terms $m^{2}(\alpha )$ leads to a non-usual renormalization of
the coupling constant $J.$ That is, we must choose $J=\mu ^{2}a^{2}$ where $%
\mu ^{2}$ has the dimension of a mass and is kept finite. Otherwise all the
particles except the one in the first Brillouin zone will decouple. In this
case we recover the usual phase where only the modes around zero are
relevant. It is trivial that the classical continuum limit is the
superposition of two uncoupled sub-lattices since $J=0$. This tree level
renormalization $J=\mu ^{2}a^{2}$ may appear unusual. However remember that
only the physical masses and coupling constants have to be cut-off
independent, not the bare parameters.

We also choose $K=\frac{1}{d}$ to get $Z(\alpha )=1$ in the continuum limit.
The appearance of the new minima is precursor of an antiferromagnetic
instabilities. That is, configurations with some antiferromagnetic
directions are metastable states in the paramagnetic phase.

With this choice for the couplings we will prove that our model describes a
well defined renormalizable field theory with $2^{d}$ interacting particles.

\section{The perturbation expansion}

We follow the standard procedure by computing the different 1-PI function at
the one loop level in $d=4$. As the initial action has only one field, it is
not trivial that the UV divergencies may be cancelled only by one mass and
one coupling counter term (it's easy to check the wave function
renormalization constant is $\delta Z=0$ at the one loop order).

We start with: 
\begin{equation}
\Gamma ^{2}(k,-k)=G^{-1}(k)+\frac{g}{2}\int_{\mathcal{B}}G(p)=G_{\alpha
}^{-1}(\widetilde{k})+\frac{g}{2}\sum_{\alpha }\int_{\mathcal{B}%
_{1}}G_{\alpha }(\widetilde{p})
\end{equation}
with $k=$ $\widetilde{k}+P(\alpha ),$ and $\widetilde{k}$ $\in \mathcal{B}%
_{1}$. As usual we replace the bare coupling by the renormalized one. The
physical renormalized mass is: 
\begin{equation}
m^{2}=\Gamma ^{2}(0)=m^{2}+\delta m^{2}+\frac{g}{2}\sum_{\alpha }\int_{%
\mathcal{B}_{1}}G_{\alpha }(\widetilde{p})
\end{equation}
which defines the mass counter term, and make the two point 1-PI function
finite. With this choice the other physical masses of the different
particles are define unambiguously by: 
\begin{equation}
m^{2}(\alpha )=\Gamma ^{2}(P(\alpha ))=m^{2}+\mu ^{2}d\sum_{\mu
=1}^{d}n_{\mu }(\alpha )
\end{equation}
The renormalization of the coupling constant is more involved. Consider:

\begin{eqnarray}
\Gamma ^{4}(k_{1},k_{2},k_{3},k_{4}) &=&g+\delta g-\frac{g^{2}}{2}\int_{%
\mathcal{B}}G(p)G(k_{1}+k_{2}+p)+ Perm \nonumber \\
&=&g+\delta g-\frac{g^{2}}{2}\sum_{\alpha }\int_{\mathcal{B}_{1}}G_{\alpha }(%
\widetilde{p})G_{\alpha }(k_{1}+k_{2}+\widetilde{p})+ Perm\nonumber \\
\end{eqnarray}
The renormalized coupling constant is defined as usual as $%
g=\lim\limits_{a\rightarrow 0}\Gamma ^{4}(0)$ so that the counter term is: 
\begin{eqnarray}
\delta g &=&\frac{3g^{2}}{2}\sum_{\alpha }\lim\limits_{a\rightarrow 0}\int_{%
\mathcal{B}_{1}}G_{\alpha }(\widetilde{p})^{2}  \nonumber \\
&=&\frac{3g^{2}}{2}\left( \sum_{\alpha }(\frac{1}{16\pi ^{2}}\ln \frac{%
\Lambda ^{2}}{m^{2}(\alpha )}-1)+F\right)  \label{ct}
\end{eqnarray}
where $F$ is a finite part due to the lattice structure, independent of $%
m^{2}(\alpha )$ (for a detailed analysis of such integrals, see \cite{af1}).
It's clear that when all the external momenta belong to $P(16),$ we have $%
\Gamma ^{4}(P(16))=\Gamma ^{4}(0)=g$. In the other cases, it is not yet
clear that the unique counter term (\ref{ct}) remove the UV divergencies. In
fact for the renormalization of the other coupling constant we have to
compute integrals of the following form where $\overline{\alpha }$ is an
implicit function of $\alpha $: 
\begin{eqnarray}
\sum_{\alpha }\lim\limits_{a\rightarrow 0}\int_{\mathcal{B}_{1}}G_{\alpha }(%
\widetilde{p})G_{\overline{\alpha }}(\widetilde{p}) &=&\sum_{\alpha }\int_{%
\mathcal{B}_{1}}\frac{dp}{(p^{2}+m^{2}(\alpha ))(p^{2}+m^{2}(\overline{%
\alpha }))}+F  \nonumber \\
&=&\sum_{\alpha }\frac{1}{16\pi ^{2}}\left( \ln \frac{\Lambda ^{2}}{%
m^{2}(\alpha )}-\frac{m^{2}(\overline{\alpha })}{m^{2}(\alpha )-m^{2}(%
\overline{\alpha })}\ln \frac{m^{2}(\alpha )}{m^{2}(\overline{\alpha })}%
\right) +F \nonumber \\
\end{eqnarray}
where the finite term $F$ is the same as above \cite{af1}. Then it is clear
that for every external momenta at the edge of the Brillouin zone, the
choice of the counter term (\ref{ct}) will remove the UV divergencies. As an
example consider the following vertex function: 
\begin{eqnarray}
\Gamma ^{4}(P(16),P(16),0,0) &=&g+\delta g-\frac{2g^{2}}{2}\int_{\mathcal{B}%
}G(p)G(P(16)+p)-\frac{g^{2}}{2}\int_{\mathcal{B}}G(p)G(-p)  \nonumber \\
&=&g+\delta g-\frac{2g^{2}}{2}\sum_{\alpha }\int_{\mathcal{B}_{1}}G_{\alpha
}(\widetilde{p})G_{17-\alpha }(\widetilde{p})-\frac{g^{2}}{2}\sum_{\alpha
}\int_{\mathcal{B}_{1}}G_{\alpha }(\widetilde{p})^{2}\nonumber \\
\end{eqnarray}
Or 
\begin{eqnarray}
\sum_{\alpha }\lim\limits_{a\rightarrow 0}\int_{\mathcal{B}_{1}}G_{\alpha }(%
\widetilde{p})G_{17-\alpha }(\widetilde{p}) &=&\sum_{\alpha
}\lim\limits_{a\rightarrow 0}\int_{\mathcal{B}_{1}}G_{\alpha }(\widetilde{p}%
)^{2}\nonumber \\
&&-\sum_{\alpha }\frac{1}{16\pi ^{2}}\frac{m^{2}(17-\alpha )}{%
m^{2}(\alpha )-m^{2}(17-\alpha )}\ln \frac{m^{2}(\alpha )}{m^{2}(17-\alpha )}
\nonumber \\
\end{eqnarray}
and finally: 
\begin{eqnarray}
\lim\limits_{a\rightarrow 0}\Gamma ^{4}(P(16),P(16),0,0) &=&g+\delta g-\frac{%
3g^{2}}{2}\sum_{\alpha }\lim\limits_{a\rightarrow 0}\int_{\mathcal{B}%
_{1}}G_{\alpha }(\widetilde{p})^{2}  \nonumber \\
&&-\sum_{\alpha }\frac{1}{16\pi ^{2}}\frac{m^{2}(17-\alpha )}{m^{2}(\alpha
)-m^{2}(17-\alpha )}\ln \frac{m^{2}(\alpha )}{m^{2}(17-\alpha )}\nonumber \\
\end{eqnarray}
which with the help of (\ref{ct}) defines another renormalized coupling
constant.

\section{The Beta function}

We deduce the following beta function from the choice of the counter term of
the coupling constant: 
\begin{equation}
\beta (g)=\left. m\frac{\partial g}{\partial m}\right| _{\lambda
_{0},\Lambda }=6g^{2}\sum_{\alpha }\frac{m^{2}}{m^{2}(\alpha )}
\end{equation}
which gives the flow for the coupling constant: 
\begin{equation}
g(\Lambda )=\frac{\lambda (\Lambda _{0})}{1-\frac{3\lambda }{16\pi ^{2}}%
\sum\limits_{\alpha }(\frac{m^{2}}{m^{2}(\alpha )})\ln \frac{\Lambda }{%
\Lambda _{0}}}
\end{equation}
Remark that the sign of the beta function may be negative if $\mu ^{2}$ $<0$%
. But a careful look shows that this happens only when $m^{2}<-16\mu ^{2}$
where the trivial vacuum is instable against the antiferromagnetic one. So
we must start again the computation by considering fluctuations around this
antiferromagnetic vacuum. In this case we found the same beta function. It
is then clear that our theory will be trivial, or in other words the
coupling constant is a non-renormalizable one, as for each scalar theory in $%
d=4$.

\section{Effective potential}

To deduce the universality class of the model the simplest way is to
introduce a formalism allowing us to compute the one loop effective
potential. It is defined as the generator function for the 1PI function as:

\begin{equation}
V_{eff}(\Phi )=\sum_{n=0}^{\infty }\frac{1}{n!}\sum_{\alpha _{1},...,\alpha
_{n}}\Phi _{\alpha _{1}}...\Phi _{\alpha _{n}}\Gamma ^{(n)}(P(\alpha
_{1}),...,P(\alpha _{n}))
\end{equation}
where we have introduced a 16-components field $\Phi _{\alpha },$ in such a
manner that the $\alpha $-th component will be responsible of the
excitations in $B_{\alpha }.$ In this manner each of these excitations is
now interpreted as low energy excitations of different fields. Thus the
Feynman rules are those of a 16-component field with the matrix propagator $%
G $ where $G_{\alpha ,\beta }(p)=\delta _{\alpha ,\beta }G(P(\alpha )+p),$
and each external line with $p=0$ is represented by the insertion of the
matrix:

\begin{equation}
\overline{\Phi }=\sum\limits_{\alpha =1}^{2^{d}}\gamma ^{\alpha }\Phi
_{\alpha }
\end{equation}
where:

\begin{equation}
\gamma^\alpha_{\rho,\sigma}=\prod_{\mu=1}^d
\delta_{\sigma_\mu+\alpha_\mu-\rho_\mu(mod2),0}
\end{equation}
takes care of the change of particle type at each vertex. Then for example
at one loop: 
\begin{eqnarray}
\frac{1}{4!}\sum_{\alpha _{1},...,\alpha _{4}}\Phi _{\alpha _{1}}...\Phi
_{\alpha _{4}}\Gamma ^{(4)}(P(\alpha _{1}),...,P(\alpha _{4}) &=&\frac{g^{2}%
}{4}\sum_{\alpha _{1},...,\alpha _{4}}\Phi _{\alpha _{1}}...\Phi _{\alpha
_{4}}\int_{\mathcal{B}_{1}}dpTr[G(p)\gamma ^{\alpha _{1}}  \nonumber \\
&&G(p)\gamma ^{\alpha _{2}}G(p)\gamma ^{\alpha _{3}}G(p)\gamma ^{\alpha
_{4}}]  \nonumber \\
&=&\frac{g^{2}}{4}\sum_{\alpha _{1},...,\alpha _{4}}\int_{\mathcal{B}%
_{1}}dpTr[G(p)\overline{\Phi }G(p)\overline{\Phi }G(p)\overline{\Phi }G(p)%
\overline{\Phi }]\nonumber \\
\end{eqnarray}
Taking advantage of the matrix formalism introduced above we obtain:

\begin{equation}
V_{eff}^{(1)}(\Phi )=\frac{1}{2}\int\nolimits_{p\le \frac{\pi }{2a}}%
\frac{d^{d}p}{(2\pi )^{d}}tr\ln \left( G^{-1}+\frac{g}{2}\overline{%
\Phi }^{2}\right)
\end{equation}
the tree level -part of the effective potential is: 
\begin{equation}
V^{(0)}(\Phi )=\sum_{\alpha }(G_{\alpha }^{-1}+\delta m^{2})\Phi _{\alpha
}^{2}+\frac{g+\delta g}{4!}(\sum_{\alpha }\Phi _{\alpha }^{4}+3\sum_{\alpha
,\beta }\Phi _{\alpha }^{2}\Phi _{\beta }^{2})
\end{equation}
It is now easy to check that the model defined by the action (\ref{action})
lies in the same universality class as a $16$ components scalar field theory
defined with only one mass counter term and one coupling constant counter
term whose Lagrangian is: 
\begin{equation}
L=\frac{1}{2}\sum_{\alpha }\partial _{\mu }\varphi _{\alpha }\partial ^{\mu
}\varphi _{\alpha }+\sum_{\alpha }\frac{m_{\alpha }^{2}+\delta m^{2}}{2}%
\varphi _{\alpha }^{2}+\frac{g+\delta g}{4!}(\sum_{\alpha }\varphi _{\alpha
}^{4}+3\sum_{\alpha ,\beta }\varphi _{\alpha }^{2}\varphi _{\beta }^{2})
\label{model2}
\end{equation}
Nevertheless an important difference is that the masses of the particles in
the model (\ref{action}) are related to each other and are then not
arbitrary as in the model (\ref{model2}).

Note that it is possible to remove most of the modes in the continuum limit
by adding a diagonal interaction. In this case the theory describes a low
energy dynamic of the two component scalar theory \cite{af1}.

\smallskip

The analysis of the present paper may be applied to other models too. For
example consider the following variant of the XY model defined by the
action: 
\[
S=\frac{1}{2T}\sum_{i}\sum_{\mu }\left\{ \alpha \cos (\theta _{i}-\theta
_{i+\mu })+\gamma \cos (\theta _{i}-\theta _{i+2\mu })\right\} 
\]
This theory will describe the usual excitations, that is the spin waves,
the vortex as well as the doubling of the modes (similar to rotons
excitations). It would be interesting to study the influences of theses
modes on the phase transitions.

\section{Conclusion}

We studied a one component scalar field theory in a d dimensional lattice
with next-nearest-neighbor interaction at the one loop level. One can
identify $2^{d}$ particle like excitation, and then eliminate the one-loop
divergencies by an appropriate fine tuning of the bare parameters. The
resulting theory is, at low energy, equivalent to a usual renormalizable $%
2^{d}$ scalar field theory. One should emphasize that the renormalized
continuum theory exists only when the regulator is taken into account both
at the tree (through the free propagator) and the one loop levels in a
systematical manner. It is interesting to pursue this work by adding
competing interactions in order to study if they may lead to continuum
physics beyond the class of traditionally (perturbative) renormalizable
theory.

\medskip

H. M. thanks V. Branchina and J.\ Polonyi for useful discussions.

\end{document}